\documentclass[pra,twocolumn,tightenlines,showpacs]{revtex4}

\usepackage[active]{srcltx} 
\usepackage{bm,dcolumn,amsmath,graphicx}

\begin{document}
\title{Forbidden M1 and E2 transitions in monovalent atoms and ions}
\author{U. I. Safronova}
\affiliation {Physics Department, University of Nevada, Reno, Nevada 89557}
\author{M. S.  Safronova}
\affiliation {Department of Physics and Astronomy, 217 Sharp Lab,
University of Delaware, Newark, Delaware 19716}
\author{W. R. Johnson}
\affiliation{Department of Physics,  225 Nieuwland Science Hall,
University of Notre Dame, Notre Dame, IN 46556}

\date{\today}
\begin{abstract}
 We carried out a systematic high-precision relativistic study of the forbidden magnetic-dipole and electric-quadrupole
 transitions in Ca$^+$, Rb, Sr$^+$, Cs, Ba$^+$,  Fr, Ra$^+$, Ac$^{2+}$ and Th$^{3+}$. This work is motivated by the
 importance of these transitions for tests of fundamental physics and precision measurements. The relative importance of the
 relativistic, correlation, Breit correction and contributions of negative-energy states is investigated.
Recommended values of reduced matrix elements are presented together with their uncertainties. The  matrix elements
and resulting lifetimes are compared with other theoretical values and with experiment where available.
\end{abstract}

\maketitle

\section{Introduction}

Forbidden transitions have been of much interest in recent years due to their applications in
optical clocks \cite{LudBoyYe15}, tests of fundamental physics \cite{science97,BerDzuFla10,DzuSafSaf15,PruRamPor15,VerWanGir11,AubBehCol13} and  quantum information \cite{SchNigMon13}.
These applications require long-lived metastable states and, therefore,  knowledge of their atomic
properties including various multipolar transition rates and branching ratios. While many accurate
measurements of the electric-dipole matrix elements exist, there are much fewer precision benchmarks for the M1 and E2 transitions.

The interest in forbidden transitions is further motivated by the emergence of the highly charged ions (HCI) as potential candidates for
the development of ultra-precise atomic clocks and tests of variation of fundamental constants \cite{BerDzuFla10,BerDzuFla11,PruRamPor15,SafDzuFla14,DzuSafSaf15}. Highly charged ions with optical
transitions suitable for metrology exhibit a particularly rich variety of low-lying multipolar transitions, even including the metastable levels that
can decay only via the M3 decay channel. Until recently, the HCI proposals remained a theoretical possibility, but the first proof-of-principle
demonstration of  sympathetic cooling of Ar$^{13+}$ with laser cooled Be$^+$
\cite{SchVerSch15} paved the way  toward the experimental realization of the HCI clock proposals.
The experimental work toward these new applications of HCIs has already started \cite{WinCreBek15} and reliable predictions of transition properties are
urgently needed. While this paper deals with ions of lower degree of ionization, up to Th$^{3+}$, the general conclusions concerning the computational accuracy and the importance of various contributions are also applicable for HCI  with a few valence electrons.

We consider examples of the forbidden transitions in Rb, Cs and Fr alkali-metal atoms and
monovalent Ca$^+$, Sr$^+$, Ba$^+$, Ra$^+$, Ac$^{2+}$ and Th$^{3+}$ ions with similar electronic structure owing to their particular interest in the applications described above as well as the  availability of some
experimental measurements. M1 transitions in Rb, Cs, Ba$^+$, Fr and Fr-like ions  are
of particular interest due to studies of parity violation  \cite{SheAndNag08,VerWanGir11,AubBehCol13,DzuFlaRob12,RobDzuFla13}.
The M1 and E2 transitions in Rb, Cs, Ba$^+$, Yb$^+$, Ra$^+$, Ac$^{2+}$ and Th$^{3+}$ ions were recently studied by
\citet{GosDzuFla13}, raising the issue of the accuracy of the M1 transition matrix elements.

The goals of parity violation studies with heavy atoms are to test the Standard Model of particle physics and to study the weak interaction inside the nucleus. In addition, atomic parity violation is  uniquely
sensitive to possible ``dark forces'' which are motivated
by the intriguing possibility of a ``dark sector'' extension
to the Standard Model  \cite{DavLeeMar14}.

The most accurate, to 0.3\%, atomic parity violation measurement was carried out in $6s-7s$ transition is Cs \cite{science97}.
The analysis of this experiment in the terms of a comparison with the standard model, which required a theoretical calculation
of the parity-violating amplitude, was carried out in \cite{PorBelDer09,DzuBerFla12}.


Here, we carry out the calculations of the E2 and M1 matrix elements for monovalent atoms and ions using a form-independent many-body perturbation theory and a relativistic linearized
coupled-cluster methods. Previous calculations of the M1 transitions, \cite{JL:76,DF:88,CH:91}, generally assumed that there are no significant corrections beyond the random-phase approximation (RPA).  Both of the methods employed in this work allow us to include corrections beyond RPA.  We find that the corrections beyond RPA are large enough to modify the results by a factor of two or more for heavier systems. We also find very strong cancelations of the various corrections beyond RPA for the $s-s$ (but not the $s-d$) transitions, causing numerical problems in the calculations, associated with the incomplete cancelation of large contributions.  We found a way to resolve these problem by using a form-independent perturbation theory as described in the paper.  We have also considered the contributions of other effects on the M1 transitions, including the two-body Breit and negative energy state contributions. We have included the study E2 transitions due to their interest of atomic clock and quantum information applications as well as availability of the experimental lifetimes for benchmark tests of the theory.
 We have conducted a systematic study of our theoretical uncertainties for the E2 matrix elements to provide recommended values for these quantities and compare them with the experimental and other theoretical values.

We start with a review of previous experimental and theoretical studies of the E2 transitions for the systems of interest.
The $s-d$ E2 transition in monovalent ions are used in clock and quantum information applications and
lifetimes of $nd$ states have been the subject of numerous studies described below.

\section{Summary of prior results for the E2 transitions}

\textbf{\textit{\textbf{Ca$^+ \,-$ \,}}}
Lifetime measurements of the metastable $3d$ levels of
Ca$^+$ were reported by Knoop {\it et al.\/} \cite{ca-95-expt} using the Ca$^+$ ions stored in a Paul
trap. The natural
lifetimes were determined to be $1111(46)$~ms
and $994(38)$~ms, for the $3d_{3/2}$ and $3d_{5/2}$ states, respectively, in agreement with
previous experiments. An improved measurement  of the $3d_{5/2}$ lifetime, 1168(7)~ms, was carried out by Barton
{\it et al.\/} \cite{ca-00-expt} using quantum jumps of a single
cold calcium ion in a linear Paul trap.
An experimental and theoretical study of the $3d$
lifetimes was reported by Kreuter {\it et al.\/}
\cite{ca-05-expt}. This work introduced a measurement technique based
on a high-efficiency quantum state detection after coherent
excitation to the $3d_{5/2}$  state or incoherent shelving in
the $3d_{3/2}$ state and subsequent free, unperturbed
spontaneous decay, yielding the value of  1168(9)~ms, in agreement with  the value reported in Ref.~\cite{ca-00-expt}.
The lifetime of the $3d_{3/2}$ state, 1176(11)~ms,
was measured with a single ion,
improving the statistical uncertainty of previous best result by a
factor of four. The experimental lifetimes were found to be in excellent agreement with the  high-precision
 {\it ab initio} all-order calculations,  [$\tau(3d_{3/2})$= 1196(11)~ms and
$\tau(3d_{5/2})$= 1165(11)~ms], reported in the same work \cite{ca-05-expt}.
Sahoo {\it et al.\/} \cite{ba-sr-ca-06} used the relativistic
coupled-cluster theory to  calculate the $3d$ lifetimes. A large-scale study of the Ca$^+$ properties, motivated by the
development of an atomic clock based on the $4s-3d_{5/2}$ transition in a Ca$^+$ single ion, was carried out in \cite{SafSaf11}. It
included the calculation of the blackbody radiation shift of the clock transition, multipole polarizabilities, oscillator strengths, lifetimes,
hyperfine constants and excitation energies.

\textbf{\textit{\textbf{Sr$^+ \,-$ \,}}}
A lifetime measurement of the metastable $4d_{3/2}$ level in
Sr$^+$ was carried out by
Mannervik {\it et al.\/} \cite{sr-prl-99} using optical pumping of a stored ion beam. Collinear laser
excitation in the storage ring transferred the main part of the
ion beam into the metastable $4d_{3/2}$ level. Subsequent
observation of the forbidden electric quadrupole transition to the
ground state yielded information about the radiative lifetime of
the metastable state,
435(4)~ms.
The lifetimes of the $4d$ levels were determined both
experimentally and theoretically by  Biemont {\it et al.\/}
\cite{sr-00-biemont}. The experiment was performed at an ion
storage ring utilizing collinear laser excitation. The calculation
was performed by the Hartree-Fock method including relativistic
effects and core polarization. The $4d_{5/2}$ lifetime was measured to be 390(1.6)~ms with a single laser-cooled,
trapped ion by Letchumanan {\it et
al.\/} \cite{sr-pra-05} using Dehmelt''s electron shelving method
to monitor the ion's electronic state.
Sahoo {\it et al.\/} \cite{ba-sr-ca-06} used the relativistic
coupled-cluster theory to calculate the $4d$ lifetimes.
A systematic  study of Sr$^+$ atomic properties was carried out in \cite{sr-10-saf}
motivated by the development of the Sr$^+$ clock and the need for the evaluation of the
blackbody radiation shift of the clock transitions.
 Safronova \cite{sr-10-saf} used the relativistic linearized
 coupled-cluster approach, which included single, double and partial triple excitations, to obtain $441(3)$~ms and
$394(2)$~ms for the lifetimes of the $4d_{3/2}$ and $4d_{5/2}$  states, respectively, in excellent agreement with
the experimental values \cite{sr-00-biemont,sr-pra-05}.

\begin{table*}
\caption{\label{tab-AcIII} Recommended values of the reduced
electric-quadrupole matrix elements in atomic units.
Dirac-Fock DF, third-order many-body perturbation theory MBPT3 and
all-order SD are listed. Final recommended values  are given
in the ``Final" column.
The relative uncertainties of the final values are given in
percent. The rows labeled NBR and BR contain results excluding and
including the Breit interaction, respectively.  Absolute values
are given.}
\begin{ruledtabular}
\begin{tabular}{lllrrrrrrrrrrrr}
\multicolumn{1}{c}{}&
\multicolumn{1}{c}{}&
\multicolumn{1}{c}{Transition}&
\multicolumn{1}{c}{DF}&
\multicolumn{1}{c}{MBPT3}&
\multicolumn{1}{c}{SD}&
\multicolumn{1}{c}{Final}&
\multicolumn{1}{c}{(\%)}&
\multicolumn{1}{c}{Transition}&
\multicolumn{1}{c}{DF}&
\multicolumn{1}{c}{MBPT3}&
\multicolumn{1}{c}{SD}&
\multicolumn{1}{c}{Final}&
\multicolumn{1}{c}{(\%)}\\
\hline
Th$^{3+}$&NBr &$7s - 6d_{3/2}$&      7.781&     6.918  &   7.063 &     7.110 &     0.66& $7s - 6d_{5/2}$&   10.008&     8.986 &   9.153  & 9.211& 0.64\\
         &Br  &$7s - 6d_{3/2}$&      7.781&     6.917  &   7.062 &     7.109 &     0.66& $7s - 6d_{5/2}$&   10.002&     8.979 &   9.145  & 9.204&  0.64\\[0.2pc]
Ac$^{2+}$&NBr &$7s - 6d_{3/2}$&     10.682&     9.218  &   9.515 &     9.585 &     0.28& $7s - 6d_{5/2}$&   13.655&    11.956 &  12.281  &12.366&  0.22\\
         & Br &$7s - 6d_{3/2}$&     10.679&     9.216  &   9.512 &     9.585 &     0.25& $7s - 6d_{5/2}$&   13.644&    11.944 &  12.270  &12.362&  0.15\\[0.2pc]
Ra$^{+}$ &NBr &$7s - 6d_{3/2}$&     17.263&    13.744  &  14.587 &    14.736 &     0.81& $7s - 6d_{5/2}$&   21.771&    17.802 &  18.689  &18.859&  0.70\\
         & Br &$7s - 6d_{3/2}$&     17.252&    13.734  &  14.578 &    14.737 &     0.74& $7s - 6d_{5/2}$&   21.749&    17.778 &  18.667  &18.855&  0.61\\[0.2pc]
Fr       &NBr &$7s - 6d_{3/2}$&     43.096&    30.292  &  31.976 &    33.427 &     0.58& $7s - 6d_{5/2}$&   52.740&    37.632 &  40.017  &41.582&  0.43\\
         & Br &$7s - 6d_{3/2}$&     43.092&    30.241  &  31.937 &    33.431 &     0.59& $7s - 6d_{5/2}$&   52.729&    37.567 &  39.963  &41.582&  0.54\\[0.2pc]
Cs       &NBr &$6s - 5d_{3/2}$&     43.846&    30.815  &  31.548 &    33.612 &     0.83& $6s - 5d_{5/2}$&   53.712&    38.087 &  39.147  &41.464&  0.57\\
         & Br &$6s - 5d_{3/2}$&     43.830&    30.763  &  31.505 &    33.620 &     1.01& $6s - 5d_{5/2}$&   53.686&    38.013 &  39.082  &41.515&  0.85\\[0.2pc]
Ba$^{+}$ &NBr &$6s - 5d_{3/2}$&     14.763&    11.821  &  12.498 &    12.627 &     0.90& $6s - 5d_{5/2}$&   18.384&    14.863 &  15.651  &15.809&  0.85\\
         &Br  &$6s - 5d_{3/2}$&     14.753&    11.813  &  12.489 &    12.627 &     0.83& $6s - 5d_{5/2}$&   18.362&    14.844 &  15.632  &15.800&  0.79\\[0.2pc]
Rb       &NBr &$5s - 4d_{3/2}$&     38.896&    31.793  &  32.444 &    32.943 &     0.42& $5s - 4d_{5/2}$&   47.636&    38.945 &  39.755  &40.367&  0.42\\
         &Br  &$5s - 4d_{3/2}$&     38.901&    31.788  &  32.156 &    32.883 &     0.24& $5s - 4d_{5/2}$&   47.642&    38.938 &  39.414  &40.295&  0.24\\[0.2pc]
Sr$^{+}$ &NBr &$5s - 4d_{3/2}$&     12.968&    10.588  &  11.010 &    11.133 &     0.35& $5s - 4d_{5/2}$&   15.972&    13.100 &  13.602  &13.747&  0.37\\
         &Br  &$5s - 4d_{3/2}$&     12.961&    10.581  &  11.003 &    11.133 &     0.30& $5s - 4d_{5/2}$&   15.957&    13.088 &  13.588  &13.745&  0.29\\[0.2pc]
Ca$^{+}$ &NBr &$4s - 3d_{3/2}$&      9.767&     7.407  &   7.788 &     7.945 &     0.48& $4s - 3d_{5/2}$&   11.978&     9.099 &   9.561  & 9.750&  0.47\\
         &Br  &$4s - 3d_{3/2}$&      9.761&     7.401  &   7.782 &     7.945 &     0.47& $4s - 3d_{5/2}$&   11.967&     9.088 &   9.552  & 9.750&  0.47 \\[0.2pc]
 \end{tabular}
 \end{ruledtabular}
\end{table*}

\textbf{\textit{\textbf{Ba$^+ \,-$ \,}}}
Lifetimes of the $5d$ states of Ba$^+$ are much longer than the corresponding
values in Ca$^+$ and Sr$^+$, making their accurate measurement particulary difficult.
A single Ba$^+$ atom was confined in a radio-frequency ion trap
and cooled by near-resonant laser light by Madej and  Sankey
\cite{ba-5d-90-expt}. A measurement of quantum-jump distributions
together with careful measurements of the absolute partial
pressures of all residual gas species enabled accurate
measurements of the quenched $5d_{5/2}$ lifetime as a
function of quenching gas pressure, 34.5$\pm$3.5~s \cite{ba-5d-90-expt}.
 The measurement of the Ba$^+$  $5d_{3/2}$ lifetime was
carried out  by Nagourney and  Dehmelt \cite{ba-prl-97}  using a single trapped Ba$^+$ ion in ultrahigh vacuum.
 The colisional quenching was found insignificant in
the experiment, but there were indications of a non-negligible
fine-structure mixing effect \cite{ba-prl-97}.
The resulting value of 79.8$\pm$4.6~s resolved the discrepancy existing at that time.
Laser-probing
measurements and calculations of lifetimes of the $5d$ levels were reported by
Gurell {\it et al.\/} \cite{ba-07-biemont}. The lifetimes,  $89.4\pm15.6$~s for the $5d_{3/2}$ level
and $32.0\pm4.6$~s for the $5d_{5/2}$ level, were measured
in a beam-laser experiment performed at the ion storage ring
CRYRING.
The electric-quadrupole transition amplitudes for Ba$^+$ were
evaluated by Gopakumar {\it et al.\/} \cite{ba-02}  using the
relativistic coupled-cluster method, giving  the
lifetimes of the $5d_{3/2}$ and $5d_{5/2}$ levels equal
to 81.4~s and 36.5~s, respectively. Sahoo {\it et al.\/}
\cite{ba-sr-ca-06} used the relativistic coupled-cluster theory to obtain
80.0(7)~s and 29.9(3)~s for these levels, respectively, followed by another
calculation of the same group   \cite{ba-5d-07}.
 Reduced
electric-quadrupole matrix elements were calculated using both many-body perturbation theory and the all-order
method including single, double and partial triple excitations by Safronova \cite{ba-5d-10}. The resulting lifetimes,  81.5(1.3)~s and 30.3(0.4)~s for the $5d_{3/2}$ and $5d_{5/2}$ levels, respectively,  were found to be in good agreement with the measured
values \cite{ba-07-biemont}.

\textit{\textbf{Cs \,- \,}}
While in alkaline-earth metal ions, the first $nd$ levels are metastable, this is not
the case in neutral alkali-metal atoms, where the first $np$ levels are below the $nd$ levels and the $E1$
decay of the $nd$ levels is allowed.
Clab  and Nayfen \cite{cs-6s-5d} measured the transition
probability of the electric quadrupole $6s-5d$ transition Cs,
$21\pm1.5$~s$^{-1}$, by two-photon ionization of the ground $6s$ state, using the $5d$ as an intermediate state. Previous measurements of this quantity yielded conflicting results.
 The authors of \cite{cs-6s-5d} noted that their  measurement was in agreement
with a laser absorption-fluorescence measurement and in
disagreement with the results of anomalous dispersion, emission,
and electron impact techniques.

\textit{\textbf{Fr and Fr-like ions \,- \,}}
Theoretical studies of the E2  $6d-7s$ transition rates were carried out in \cite{fr-6d-15}.
Safronova {\it et al.\/} \cite{ra-ac-07} calculated reduced matrix
elements of the E2 $6d-7s$ transitions in Fr-like Ra and Ac ions
using the relativistic linearized coupled-cluster method.

The M1 and E2 transitions in Rb, Cs, Ba$^+$, Yb$^+$, Ra$^+$, Ac$^{2+}$ and Th$^{3+}$ ions were  studied by
\citet{GosDzuFla13}.

\begin{table}
\caption{\label{tab-comp-e2} Recommended values of the reduced
electric-quadrupole
 matrix elements (in a.u.)  are compared
with experimental measurements and other theoretical values.}
\begin{ruledtabular}
\begin{tabular}{lcrrr}
\multicolumn{1}{c}{}&
\multicolumn{1}{c}{Transition}& \multicolumn{1}{c}{Present}&
\multicolumn{1}{c}{Expt.}&
\multicolumn{1}{c}{Theory}\\
\hline
Th$^{3+}$&$7s - 6d_{3/2}$& 7.110(47) &           &7.10 \cite{GosDzuFla13}\\
Ac$^{2+}$&$7s - 6d_{3/2}$&  9.585(27)&           &9.58 \cite{GosDzuFla13}\\
Ra$^+$   &$7s - 6d_{3/2}$&  14.74(12)&           &14.77 \cite{GosDzuFla13}\\
Fr       &$7s - 6d_{3/2}$&  33.43(19)&           &35.96(60)\cite{fr-6d-15}\\
         &               &           &           &33.59 \cite{GosDzuFla13}\\
Cs       &$6s - 5d_{3/2}$&   33.61(28) &34.2(1.2) \cite{cs-6s-5d}& 33.60 \cite{GosDzuFla13}\\
Ba $^+$  &$6s - 5d_{3/2}$&   12.63(11)&12(1) \cite{ba-07-biemont} &12.74 \cite{ba-5d-07}\\
         &               &            &12.76(35) \cite{ba-prl-97}     &12.63 \cite{ba-02}\\
Rb       &$5s - 4d_{3/2}$&   32.94(14)&                                  &33.42 \cite{GosDzuFla13}\\
Sr$^+$   &$5s - 4d_{3/2}$&   11.13(39)&11.21(5) \cite{sr-prl-99}    &11.33(10) \cite{ba-sr-ca-06}\\
         &               &            &11.21(5) \cite{sr-00-biemont}&11.13(4) \cite{sr-10-saf}\\
Ca$^+$   &$4s - 3d_{3/2}$&    7.94(4)&8.01(4) \cite{ca-05-expt} &7.94(4) \cite{ca-05-expt}\\
         &               &            &7.92(3) \cite{ca-00-expt}& 7.97(2) \cite{ba-sr-ca-06}
\end{tabular}
\end{ruledtabular}
\end{table}

\section{Electric-quadrupole transitions}

For electric-quadrupole transitions, we carried out all calculations using four different variants of the
linearized coupled-cluster (all-order) method. A review of the all-order method, which involves summing
series of dominant many-body perturbation terms to all orders,
is given in \cite{SafJoh08}. In the single-double (SD) all-order approach, single and double excitations of the
Dirac-Fock orbitals are included and the SD state vector
of a monovalent atom in state $v$:
\begin{multline}
\lefteqn{ |\Psi_v \rangle = \left[ 1 + \sum_{ma} \, \rho_{ma}
a^\dagger_m a_a + \frac{1}{2} \sum_{mnab} \rho_{mnab} a^\dagger_m
a^\dagger_n a_b a_a +
 \right. }  \\
 + \left. \sum_{m \neq v} \rho_{mv} a^\dagger_m a_v + \sum_{mna}
\rho_{mnva} a^\dagger_m a^\dagger_n a_a a_v \right]| \Psi_v^{(0)}\rangle
, \label{eq1}
\end{multline}
where $|\Psi_v^{(0)}\rangle$ is the lowest-order atomic state vector and $a^\dagger_i$,  $a_j$ are creation and annihilation
operators. The quantities
 $\rho_{ma}$, $\rho_{mv}$ are single-excitation coefficients  for  core and valence electrons
 and  $\rho_{mnab}$ and $\rho_{mnva}$ are double-excitation coefficients for core and valence electrons, respectively.
In Eq.~(\ref{eq1}), the indices $m$ and $n$ range over all possible virtual states while
 indices $a$ and $b$ range over all occupied core states.
 The single, double, partial-triple (SDpT) method also includes classes of the triple
excitations.

In either SD or SDpT all-order method, the matrix elements of any one-body operators, such as M1 and E2,
\begin{equation}
  Z = \sum_{ij} z_{ij} a^\dagger_i a_j,
\end{equation}
are
obtained as
\begin{equation}
Z_{wv}=\frac{\langle \Psi_w |Z| \Psi_v \rangle}{\sqrt{\langle \Psi_v
| \Psi_v \rangle \langle \Psi_w | \Psi_w \rangle}}, \label{eqr}
\end{equation}
where $|\Psi_v\rangle$  and $|\Psi_w\rangle$ are given by the
expansion (\ref{eq1}). The numerator of
Eq.~(\ref{eqr}) consists of the sum of the lowest-order DF matrix element
$z_{wv}$ and twenty other terms that are linear or quadratic
functions of the excitation coefficients $\rho_{mv}$, $\rho_{ma}$, $\rho_{mnva}$,
and $\rho_{mnab}$.

\begin{table}
\caption{\label{lifetimes} Lifetimes ${\tau}$ of the $nd$
states in  Ca$^+$, Sr$^+$, Ba$^+$, Ra$^+$ and Ac$^{2+}$ in seconds. }
\begin{ruledtabular}
\begin{tabular}{lllll}
\multicolumn{1}{c}{Ion}&
\multicolumn{1}{c}{State}&
\multicolumn{1}{c}{Present}&
\multicolumn{1}{c}{Theory}&
\multicolumn{1}{c}{Experiment}\\
\hline
Ca$^+$&$3d_{3/2}$& 1.194(11)&0.98   \cite{ca-pra-88-kim}    & 1.111(46) \cite{ca-95-expt}    \\
 &&&1.271   \cite{nd-wrj-91}            & 1.17(5) \cite{ca-3d-expt-99a}  \\
 &&&1.16   \cite{ca-pra-92-froese}      & 1.20(1)  \cite{ca-00-expt}  \\
 &&&1.080          \cite{ca-liaw-95}    & 1.176(11) \cite{ca-05-expt}\\
 &&&1.196(11)\cite{ca-05-expt} &                                                \\
 &&&1.185(7)
 \cite{ba-sr-ca-06}     &                                                \\[0.5pc]

 Sr$^+$&$4d_{3/2}$&0.437(14) &0.454           \cite{nd-wrj-91}    &  0.435(4)\cite{sr-prl-99}\\
 &&& 0.422           \cite{sr-00-biemont}    &  0.435(4)\cite{sr-00-biemont}\\
 &&& 0.426(7)\cite{ba-sr-ca-06}    &  0.455(29)\cite{sr-00-biemont}\\
 &&&  0.441(3)\cite{mar-sr}& \\
 [0.5pc]

 Ba$^+$&$5d_{3/2}$&81.4(1.4)& 83.7   \cite{nd-wrj-91}             &  79.8(4.6) \cite{ba-prl-97}\\
 &&& 81.5   \cite{dzuba01}               & 89(16) \cite{ba-07-biemont}\\
 &&& 81.4  \cite{ba-02}           &                                             \\
 &&& 80.1(7)\cite{ba-sr-ca-06}    &\\
 &&& 82.0  \cite{ba-07-biemont}           &\\
 &&& 81.5(1.2)\cite{ba-5d-saf-08}     &\\
&&& 84.5\cite{RobDzuFla13}     &\\[0.5pc]

 Ra$^+$ &$ 6d_{3/2}$  &  0.6382(94)&0.638(10)\cite{pra-safr-09}&\\
 &&& 0.627(4)\cite{pra-sahoo}&\\
&&& 0.642\cite{dzuba01}&\\
&&& 0.642\cite{RobDzuFla13}&\\[0.5pc]

 Ac$^{2+}$&$ 6d_{3/2}$  &   1.171(6)$\times10^6$& 1.19 $\times10^6$\cite{RobDzuFla13} &\\[1ex]
\hline\\
Ca$^+$&$3d_{5/2}$&1.163(11)&0.95 \cite{ca-pra-88-kim}             & 0.994(38) \cite{ca-95-expt}     \\
&&&1.236 \cite{nd-wrj-91}                & 1.064(17)  \cite{ca-3d-expt-96} \\
&&&1.14 \cite{ca-pra-92-froese}          & 0.969(21)  \cite{ca-3d-expt-97} \\
&&&1.045        \cite{ca-liaw-95}        & 1.09(5)  \cite{ca-3d-expt-99a} \\
&&&1.165(11) \cite{ca-05-expt}           & 1.100(18)  \cite{ca-3d-expt-99} \\
&&&1.110(9) \cite{ba-sr-ca-06}           & 1.168(7)  \cite{ca-00-expt} \\
&&&                                      & 1.168(9)  \cite{ca-05-expt} \\
&&&                                      & 1.174(10)  \cite{3d-expt-15} \\[0.5pc]

Sr$^+$&$4d_{5/2}$& 0.3945(22)& 0.405          \cite{nd-wrj-91} &  0.372(25)   \cite{sr-4d-expt-90}\\
&&& 0.384          \cite{sr-00-biemont} &  0.408(22)   \cite{sr-00-biemont}\\
&&& 0.357(12)\cite{ba-sr-ca-06} &  0.3908(16) \cite{sr-pra-05}\\
&&&  0.394(3)\cite{mar-sr}& \\
[0.5pc]

Ba$^+$&$5d_{5/2}$&30.34(48)&  37.2 \cite{nd-wrj-91}             &  32(5) \cite{ba-5d-expt-86}\\
 &&& 30.3 \cite{dzuba01}               &  34.5(3.5) \cite{ba-5d-90-expt}\\
 &&& 36.5 \cite{ba-02}          & 32.0(4.6) \cite{ba-07-biemont}\\
 &&&29.9(3) \cite{ba-sr-ca-06} & 31.2(09)\cite{5d-expt-14}\\
 &&& 31.6 \cite{ba-07-biemont}          &\\
 &&& 30.4(4) \cite{ba-5d-saf-08}  &\\[0.5pc]

 Ra$^+$ &$ 6d_{5/2}$  &  0.3028(37)&0.303(4)\cite{pra-safr-09}&0.232(4)\cite{rad-pra}\\
 &&& 0.297(4)\cite{pra-sahoo}&0.232(4)\cite{rad-cjp}\\
&&& 0.302\cite{dzuba01}&\\[0.5pc]
Ac$^{2+}$&$ 6d_{5/2}$&    2.326(34)  &&\\
\end{tabular}
\end{ruledtabular}
\end{table}

\begin{table*}
\caption{\label{tab-neg}The  M1
 matrix elements evaluated in the second-order RMBPT
approximation.
The lowest-order matrix elements
without and with  retardation, are listed in columns labeled
``DF'' and
 ``DF(Ret)''. The second
order Coulomb and Breit contributions are listed in the ``Cl'' and ``Br$_{\rm pos}$'' columns. The second-order Breit correction calculated with the
inclusion of the negative
energy (NEG) contributions is given in next column, Br$_{\rm neg}$. The final
second-order results MBPT2=Cl+Br$_{\rm pos}$+Br$_{\rm neg}$
 are listed in the last column. Units: $10^5 \mu_B$.}
\begin{ruledtabular}
\begin{tabular}{llrrrrrr}
\multicolumn{1}{c}{}&
\multicolumn{1}{c}{Transition}
& \multicolumn{1}{c}{DF}&
\multicolumn{1}{c}{DF(ret)}& \multicolumn{1}{c}{Cl}& \multicolumn{1}{c}{Br$_{\rm pos}$}&
\multicolumn{1}{c}{Br$_{\rm neg}$}&
\multicolumn{1}{c}{MBPT2}\\
\hline
Th$^{3+}$ &$ 7s -  8s_{1/2}$& 14.44 &  11.78 & -127.9 & -0.232  &  3.177&  -125.1\\
Ac$^{2+}$ &$ 7s -  8s_{1/2}$& 10.07 &  8.336 & -130.1 & -0.136  &  2.304&  -128.1\\
Ra$^{+}$  &$ 7s -  8s_{1/2}$& 6.085 &  5.141 & -129.8 &  0.142  &  1.252&  -128.5\\
Fr        &$ 7s -  8s_{1/2}$& -2.559& -2.229 &  56.82 &  -0.277 & -0.353&   56.22\\
Cs        &$ 6s -  7s_{1/2}$&  1.952&  1.631 & -5.001 &  0.236  &  0.171&  -4.617\\
Ba$^{+}$  &$ 6s -  7s_{1/2}$&  4.952&  4.021 & -12.710&  0.140  &  0.821&  -11.80\\
Rb        &$ 5s -  6s_{1/2}$&  1.824&  1.479 &  0.288 &  0.207  &  0.116&   0.588\\
Sr$^{+}$  &$ 5s -  6s_{1/2}$&  4.800&  3.784 & -2.123 &  0.057  &  0.730&  -1.390\\
Ca$^{+}$  &$ 4s -  5s_{1/2}$& -4.395& -3.308 & -0.063 & -0.014  & -0.530&  -0.557\\[0.4pc]
Th$^{3+}$ &$ 7s -  6d_{3/2}$&  1.545&  1.560 &  147.3 &  0.024  & -1.127&   146.3\\
Ac$^{2+}$ &$ 7s -  6d_{3/2}$&  1.120&  1.121 &  145.2 &  0.029  & -1.266&   144.1\\
Ra$^{+}$  &$ 7s -  6d_{3/2}$& -0.596& -0.641 & -39.46 & -0.099  &  1.380&  -38.25\\
Fr        &$ 7s -  6d_{3/2}$& -0.063&  0.074 &  26.53 &  0.159  & -0.971&   25.75\\
Cs        &$ 6s -  5d_{3/2}$&  0.094& -0.026 & -2.705 & -0.091  &  0.732&  -2.089\\
Ba$^{+}$  &$ 6s -  5d_{3/2}$& -0.551& -0.562 & -14.52 &  0.036  &  0.736&  -13.81\\
Rb        &$ 5s -  4d_{3/2}$& -0.289& -0.120 & -0.731 &  0.054  &  -0.505& -1.165\\
Sr$^{+}$  &$ 5s -  4d_{3/2}$&  0.155&  0.206 &  2.612 & -0.019  &  -0.713&  1.945\\
Ca$^{+}$  &$ 4s -  3d_{3/2}$& -0.041& -0.090 & -0.347 &  0.024  &   0.625&  0.237\\
\end{tabular}
\end{ruledtabular}
\end{table*}

The largest terms are frequently
 \begin{eqnarray}
    Z^{(a)} &=& \sum_{ma} z_{am} \tilde{\rho}_{wmva} +\sum_{ma}z_{ma} \tilde{\rho}^*_{vmwa}, \\
    Z^{(c)} &=&  \sum_{m}z_{wm} \rho_{mv} +\sum_{m}z_{mv} \rho^*_{mw}.
    \label{termc}
\end{eqnarray}
The first of these terms $Z^{(a)}$ is associated with the random-phase approximation (RPA) corrections, while the second $Z^{(c)}$ is associated  with the
Brueckner-orbital corrections; however, there is not a one-to-one correspondence to the many-body classification of
corrections to matrix elements.

Omitted higher excitations can also be estimated by the scaling procedure described in
\cite{SafJoh08}, which corrects the  $\rho_{mv}$ excitation coefficients and the corresponding terms
containing these quantities in Eq.~(\ref{eqr}), such as term (c). The scaling procedure can be applied to either SD or SDpT approximations.
The resulting values are labeled with the subscript $sc$, SD$_{\rm sc}$ and SDpT$_{\rm sc}$.
Comparing values obtained in different approximations, \textit{ab initio} SD and SDpT and scaled SD and SDpT allows us to evaluate the uncertainty of the calculations in the cases where the contributions that can be corrected by scaling  are dominant.
 We find that this condition is satisfied
for the E2 transitions considered in the present work, where term c given by Eq.~(\ref{termc}) strongly dominates.

In Table \ref{tab-AcIII}, we list our recommended values for the $s - d$ E2  reduced matrix
elements in Fr and Fr-like ions,  Cs,  Ba$^+$, Rb,
Sr$^+$ and Ca$^+$.
The absolute values
are given  in units of $ea_0^2$, where $a_0$ is the Bohr radius and $e$ is the elementary charge.
Results of first-order Dirac-Fock, third-order many-body perturbation theory and the four all-order calculations described above are listed in the columns
labeled DF, MBPT3, SD, SDpT, SD$_{\rm sc}$ and SDpT$_{\rm sc}$.
 We  also carried out the calculations using form-independent third-order many-body perturbation theory (MBPT3)
method introduced in \cite{SavJoh00,SavJoh00a}.
The lowest-order values, given in the DF column,  illustrate the size of the
correlation corrections. The difference of the MBPT3 and the all-order results
illustrates the size of the higher-order corrections beyond random-phase approximation, which are
included to all orders in MBPT3.    Final recommended values are given
in the ``Final'' column. The next
column gives the absolute uncertainties.  The evaluation of the
uncertainty of the matrix elements in this approach was described
in detail in \cite{SafSaf11rb,SafSaf12}.  The differences of the
all-order values for each transition calculated in different approximation were used to estimate uncertainty in the final
results based on the algorithm that accounted for the importance of
the specific dominant contributions.
The column labeled ``Unc. \%'' of Table~\ref{tab-AcIII} gives relative uncertainties
of the final values  in percent.  The uncertainties are small and range from 0.1\% to 1\%.

We also investigated the effect of the Breit interaction on the
E2 matrix elements. Table~\ref{tab-AcIII} lists the results calculated with and without the Breit interaction. The one-body part of the Breit interaction was included in the
construction of the finite basis set which was used in all of the all-order calculations. The two-body Breit
correction to matrix elements is small as discussed in detail in \cite{Der02}.
The Breit contribution is very small,
less than 0.01\% for all cases. The relative uncertainties given in the last
column of Table~\ref{tab-AcIII} are less than 1\%.

In Table~\ref{tab-comp-e2}, our recommended values of the reduced electric quadrupole matrix
elements are compared with recent theoretical calculations of
Ref.~\cite{GosDzuFla13}.
Most of the other theoretical and all of the experimental papers give
the results for the lifetimes of the $nd$ states of ions levels rather than the
$s - d$ matrix elements. For the $nd_{3/2}$ lifetimes $\tau$, the E2  matrix elements $Z(nd_{3/2}-n^{\prime}s)$ in a.u. may be accurately extracted
using
$$
Z=\left[ \frac{(2j+1)\lambda^{5}} {1.11995\times 10^{18} \, \, \tau} \right]^{-1/2},
$$
where $j=3/2$, $\lambda$ is the wavelength of the $ns-n^{\prime}d_{3/2}$ transition in~\AA and lifetimes $\tau$ is in seconds.
The contribution of the $ns-n^{\prime}d_{3/2}$ transitions
is negligibly small.

\begin{table*}
\caption{\label{tab-M1} Recommended values of the reduced magnetic
M1 dipole
 matrix elements in $10^5\mu_B$.
 The first-order and all-order SD  values are listed;
 Final recommended values  are given
 in the ``Final'' column. RPA includes lowest-order DF results,
 the third-order MBPT results (MBPT3) includes both DF and RPA
 results. The results are compared with DF and RPA values from
 \cite{GosDzuFla13}.}
\begin{ruledtabular}
\begin{tabular}{llrrrrrrrr}
\multicolumn{1}{c}{}&
 \multicolumn{1}{c}{}&
\multicolumn{1}{c}{Transition}&
\multicolumn{1}{c}{DF}&
\multicolumn{1}{c}{DF~\cite{GosDzuFla13}}&
\multicolumn{1}{c}{RPA}&
\multicolumn{1}{c}{RPA~\cite{GosDzuFla13}}&
\multicolumn{1}{c}{MBPT3}&
\multicolumn{1}{c}{SD}&
\multicolumn{1}{c}{Final}\\
\hline
Th$^{3+}$&No Breit &$7s - 6d_{3/2}$&  1.545&        & 214.4&        &111.9  &121.9&   121.9   \\
         &   Breit &$7s - 6d_{3/2}$&  4.394& 4.432  & 216.0& 212.2  &123.8  &123.1&   123.1   \\ [0.2pc]

Ac$^{2+}$&No Breit &$7s - 6d_{3/2}$&  1.119&        & 216.3&        &123.1  &129.6&   129.6   \\
         &   Breit &$7s - 6d_{3/2}$&  3.513& 3.510  & 214.5& 213.6  &113.2  &130.3&   130.3   \\[0.2pc]

Ra$^{+}$ &No Breit &$7s - 6d_{3/2}$&  0.596&        & 213.7&        &142.5  &138.4&   138.4   \\
         &   Breit &$7s - 6d_{3/2}$&  2.368& 2.401  & 212.8& 210.3  &142.5  &138.6&   138.6   \\[0.2pc]

Fr       &No Breit &$7s - 6d_{3/2}$&  0.063&        & 128.5&        &146.3  &125.9&   125.9   \\
         &   Breit &$7s - 6d_{3/2}$&  0.570& 0.737  & 127.9& 126.9  &145.8  &125.6&   125.6   \\[0.2pc]
Cs       &No Breit &$6s - 5d_{3/2}$&  0.094&        & 12.70&        &13.52  &13.23&   13.23   \\
         &   Breit&$ 6s - 5d_{3/2}$&  0.429& 0.566  & 12.95& 11.98  &14.05  &13.84&   13.84   \\[0.2pc]
Ba$^{+}$ &No Breit &$6s - 5d_{3/2}$&  0.551&        & 22.72&        &13.54  &15.65&   15.65   \\
         &   Breit&$6s  - 5d_{3/2}$&  2.009& 2.006  & 23.65& 22.06  &14.86  &16.94&   16.94   \\[0.2pc]
Rb       &No Breit&$5s  - 4d_{3/2}$& -0.289&        & 1.214&        &0.849  &1.553&   1.553   \\
         &   Breit&$5s  - 4d_{3/2}$&  0.017& 0.245  & 1.448& 1.019  &1.238  &2.006&   2.006   \\[0.2pc]

Sr$^{+}$ &No Breit&$5s  - 4d_{3/2}$&  0.155&        & 4.125&        &1.853  &3.380&   3.380   \\
         &   Breit&$5s  - 4d_{3/2}$&  1.463&        & 5.210&        &3.193  &4.693&   4.693   \\[0.2pc]

Ca$^{+}$ &No Breit&$4s  - 3d_{3/2}$&  0.041&        &-0.657&        &0.040  &0.817&   0.817   \\
         &   Breit&$4s  - 3d_{3/2}$&  1.203&        & 1.708&        &1.183  &2.012&   1.973   \\
 \end{tabular}
\end{ruledtabular}
\end{table*}

For the $nd_{5/2}$ states, there is an additional contribution to the lifetime
from the $nd_{5/2}-nd_{3/2}$ M1 transition.
 In light ions, Ca$^+$ and Sr$^+$, the contribution
of this M1 decay channel to the lifetime is very small, but it becomes significant for Ba$^+$, with 18\% branching ratio, i.e.
relative contribution of the M1 rate to the sum of the M1 and the E2 transition rates.
Our matrix
elements for the $6s-5d_{3/2}$ transitions in Cs is in excellent
agreement with experimental measurements given in
Ref.~\cite{cs-6s-5d}, with the theoretical prediction having much smaller uncertainty.
Our values are in agreement with the experiment for alkaline-earth ions within the uncertainties.

 Table~\ref{lifetimes} gives the comparisons of the present lifetime results with the experiment
  \cite{ca-95-expt,ca-00-expt,ca-3d-expt-96,ca-3d-expt-97,ca-3d-expt-99a,ca-05-expt,ca-3d-expt-99,ba-sr-ca-06,sr-00-biemont,ba-prl-97,ba-07-biemont,3d-expt-15}
 and with other theory \cite{ca-pra-88-kim,ca-3d-expt-99a,ca-pra-92-froese,ca-liaw-95,ca-pra-88-kim,nd-wrj-91,sr-prl-99,
sr-00-biemont,ba-sr-ca-06,ba-02,ba-5d-saf-08,mar-sr,dzuba01,RobDzuFla13,pra-safr-09,pra-sahoo}.
 No  experimental lifetimes are available for
 the $6d$ levels of the Fr-like ions.

In alkali-metal neutral atoms, $nd$ states are not metastable and E2 or M1 contributions to the lifetimes
are negligible.

\begin{table}
\caption{\label{tab-m1j} Magnetic dipole (M1) matrix elements  in
units of $10^5\mu_B$.  Relative signs of the present results are
adjusted so that final matrix elements are positive. RPA includes
lowest-order DF results; the third-order MBPT results (MBPT3)
include both DF and RPA.  The results are compared with DF and RPA
values from \cite{GosDzuFla13}}
\begin{ruledtabular}
\begin{tabular}{llrrrrrr}
\multicolumn{1}{c}{}&
 \multicolumn{1}{c}{}&
\multicolumn{1}{c}{Tran}& \multicolumn{1}{c}{DF}&
\multicolumn{1}{c}{DF~\cite{GosDzuFla13}}&
\multicolumn{1}{c}{RPA}&
\multicolumn{1}{c}{RPA~\cite{GosDzuFla13}}&
\multicolumn{1}{c}{MBPT3}\\
\hline
Th$^{3+}$&NBr &$8s -7s$& -14.44 &            &158.8 &           &  64.68  \\
         &Br  &$8s -7s$& -15.78 &   -13.23   &155.2 &  -2549    &  61.66  \\ [0.2pc]
Ac$^{2+}$&NBr &$8s -7s$& -10.07 &            &172.8 &           &  86.97  \\
         &Br  &$8s -7s$& -11.13 &   -8.911   &169.4 &  -2390    &  84.18  \\[0.2pc]
Ra$^{+}$ &NBr &$8s -7s$& -6.085 &            &185.3 &           &  112.7  \\
         &Br  &$8s -7s$& -6.935 &   -5.744   &182.0 &   185.1   &  110.0  \\[0.2pc]
Fr       &NBr &$8s -7s$& -2.559 &            &177.1 &           &  139.9  \\
         &Br  &$8s -7s$& -3.000 &   -2.49    &174.4 &   176.5   &  137.4  \\[0.2pc]
Cs       &NBr &$7s -6s$& -1.952 &            &14.22 &           &  12.45  \\
         &Br  &$7s -6s$& -2.189 &   -1.652   &13.66 &   14.13   &  11.83  \\[0.2pc]
Ba$^{+}$ &NBr &$7s -6s$& -4.952 &            &13.24 &           &  8.042  \\
         &Br  &$7s -6s$& -5.366 &   -4.050   &12.46 &   13.53   &  7.257  \\[0.2pc]
Rb       &NBr &$6s -5s$& -1.824 &            &1.004 &           &  0.859  \\
         &Br  &$6s -5s$& -1.998 &   -1.473   &0.740 &   1.216   &  0.553  \\[0.2pc]
Sr$^{+}$ &NBr &$6s -5s$&  4.800 &            &1.706 &           &  2.392  \\
         &Br  &$6s -5s$&  5.099 &            &2.099 &           &  2.828  \\[0.2pc]
Ca$^{+}$ &NBr &$5s -4s$&  4.395 &            &4.400 &           &  4.579  \\
         &Br  &$5s -4s$&  4.570 &            &4.579 &           &  4.791  \\
 \end{tabular}
\end{ruledtabular}
\end{table}

\section{Magnetic dipole matrix elements }

The M1 matrix elements for the $s-s$ and $s-d$ transitions are much more difficult to calculate accurately than the E2 ones.
For the E2 transitions, the correlation contributes at most 25\% to the  total, while for the M1 transitions the lowest-order
values are very small and the final result comes almost entirely from the correlation corrections.
The Breit interaction is more significant as well. Moreover, the negative-energy states, $\varepsilon_{i} < mc^2$, may contribute.

The influence of the negative-energy states (NES) on forbidden
magnetic-dipole $s - s$ transitions in
alkali-metal atoms was investigated by Savukov {\it et al.\/} in
Ref.~\cite{savukov-prl-99}. The NES
contributions were significant in almost all cases and, for
rubidium, reduced the transition rate by a factor of 8.
Derevianko {\it et al.\/} \cite{savukov-pra-98} derived the
leading term in an $\alpha Z$ expansion for the negative-energy
(virtual electron-positron pair) contributions to the transition
amplitudes of heliumlike ions, finding  a strong
dependence on the choice of the zeroth-order Hamiltonian, which
defines the negative-energy spectrum.
 The ratio of negative-energy contributions to
the total transition amplitudes for some nonrelativistically
forbidden transitions was shown to be of order $1/Z$. In the
particular case of the magnetic-dipole transition $3\ ^3S_{1}- 2\
^3S_1$, authors noted that neglecting of negative-energy
contributions, in an otherwise exact no-pair calculation, would
lead one to underestimate the decay rate in helium by a factor of
1.5 in calculations using a Hartree basis and by a factor of 2.9
using a Coulomb basis \cite{savukov-pra-98}.

The contribution from the negative-energy states for the M1 transitions in
Be-like ions was studied by Safronova {\it et
al.\/}\cite{saf-scr-m1}, demonstrating that the NES contribution
scales as $\alpha ^{2}Z$ for both Breit and Coulomb interactions.
The relative contribution of the NES was about
0.03\% for transitions inside the $2l2l'$ configuration space and
3\% for the $2l_{1}2l_{2}-2l_{3}3l_{4}$ transition. Authors
concluded that the NES contributions were important for the
weakest transitions in a given transition array.

 The E1, E2, M1 and M2 transitions in the nickel
isoelectronic sequence were investigated by  Hamasha {\it et
al.\/} \cite{alla-cjp-04}. The contributions from negative-energy
states were included in the second-order E1, M1, E2 and M2 matrix
elements. In second-order matrix elements, such contributions
arise explicitly from those terms in the sum over states for which
$\varepsilon_{i} < mc^2$.  The NES contributions drastically
change the second-order Breit-Coulomb matrix elements $B^{(2)}$.
However, the second-order Breit-Coulomb correction contributes
only 2-5\% to uncoupled M1 matrix elements and, as a result,
negative-energy states changed the total values of M1 matrix
elements by only a few percent \cite{alla-cjp-04}.

The contributions from negative-energy states were included in the
second-order E1, M1, E2 M2, E3 and M3 matrix elements in
\cite{safr-ni-06}.  The NES contributions to the second-order
Breit-Coulomb matrix elements for the transition from
$3d_{5/2}5d_{3/2}(1)$ state in Ni-like ions weakly increases with $Z$, however,
the relative NES contribution for this transition decreases with
$Z$ (2\% and 0.6\% for $Z$ = 40 and $Z$ = 90, respectively). Ref.~\cite{safr-ni-06} noted
 that the NES contribution for
this transition are of the same order as the positive-energy state
contribution to the second-order Breit-Coulomb matrix elements
causing severe cancelation and drastically reducing the $B^{(2)}$
values in this case.
Therefore, we  include the contributions of the NES as well as retardation corrections and correlation effects for the M1 transitions in detail in the second-order MBPT calculation.

In Table~\ref{tab-neg}, we list the  M1 magnetic matrix elements evaluated in second-order RMBPT approximation.
We employ customary units for reduced matrix elements as given in the National Institute for Science and Technology compendium on Atomic Spectroscopy \cite{NIST17}, These units are $e^2a_0^2$ for E2 transitions and $\mu_B$ for M1 transitions.


The lowest-order DF values are evaluated with the relativistic version of the M1 operator
 without retardation.
The  DF(Ret) values
include retardation. The table illustrates that the retardation corrections are particularly large for the $s-d$
transition in Rb, Cs and Fr.
 The second-order Coulomb and Breit contributions are listed in the ``Cl'' and ``Br$_{\rm pos}$'' columns. The second-order Breit correction,
which includes the negative
energy (NEG) contributions, is given in next column, Br$_{\rm neg}$. The final
second-order results MBPT2=Cl+Br$_{\rm pos}$+Br$_{\rm neg}$
 are listed in the last column.
  We find that the NES effect on the Coulomb correlation correction is negligible and can be omitted without the loss of accuracy,
 and it is not shown in the table.
The contribution of the NES to the second-order Breit correction is significant, as illustrated by the significant differences of the
$B_{\rm pos}$ and $B_{\rm neg}$ values.
However, the table clearly indicates that the Coulomb correlation correction dominates the final values and
an accurate calculation of this correction presents a
significant challenge.
As noted above, the M1 transitions between levels of different electronic configurations are extremely sensitive to the correlation correction, since the lowest-order M1 values are very small and the final
result comes almost entirely from the correlation correction.

While it was previously assumed that only RPA corrections contribute significantly to the M1 matrix elements, we find that it is not
the case for the transitions  studied in this work.

In Table \ref{tab-M1}, we list our values for the M1 $s - d$
reduced matrix
elements in units of $10^5 \mu_B$. The final results are obtained using the same all-order approach as for the
E2 matrix elements. The four variants of the all-order calculations are carried out for the M1 transitions as for the E2 transitions. We  also carried out the calculations of the M1 matrix elements using form-independent third-order many-body perturbation theory (MBPT3)
method introduced in \cite{SavJoh00,SavJoh00a}.
 The all-order values are taken as final.
We find that while using the form of the M1 operator that includes retardation  changes the DF values, its effect
on the final result is negligible at the present level of accuracy and is omitted in Table~\ref{tab-M1}.
 Results with and without the inclusion of the Breit interaction are listed, with the Breit
contribution  being more important for the M1 transitions in comparison with the E2 transitions.

  Comparing the third-order MBPT3 and RPA results demonstrates that
  corrections beyond RPA are large for all cases, in particulary Fr and Fr-like ions.
  The MBPT3 classification and formulas for such corrections, which include Brueckner-orbital (BO), structure radiation (SR), and normalization
    is given in \cite{life-tran-96}. The form-independent variant of the third-order used here
    includes further corrections due to replacement of the DF matrix elements by the ``dressed'' RPA values in all formulas. This approach is discussed in detail in \cite{SavJoh00,SavJoh00a}.
    The all-order SD calculations include all of the third-order and additional higher-order correlation corrections. The comparison of the MBPT3 and all-order SD values demonstrate that
    the fourth and higher-order contributions are significant for these M1 transitions.

    The same calculations are carried out for the $s-s$ transitions. The results are presented in Table~\ref{tab-m1j}, where DF, RPA and the MBPT3 final values are listed.
      Our M1 values are compared with the  theoretical results from
     Ref.~\cite{GosDzuFla13} obtained in the DF and RPA approximations.
     Negative-energy and retardation corrections are omitted, these contributions are smaller than the uncertainty in the correlation corrections as demonstrated in Table \ref{tab-neg}. DF energies are used to define $\omega$ in all RPA calculations.  The MBPT3 values are taken as final.

    We identified two issues in the calculations of these matrix elements. First, we find that there are significant numerical instabilities in Dirac-Fock computations of the M1 $s-s$ matrix elements (the effect is small for the $s-d$ case). The DF codes used for generation of the several low-level orbitals do not usually orthonormalize the resulting wave functions, since it is done by subsequent basis set codes. In the relativistic case one expects the accuracy of the M1 radial matrix element to be limited by the size of the overlap matrix integral $(g_v g_w + f_v f_w) dr$, where $g$ and $f$ are the large and small components of the wave function and $v$ and $w$ indicate initial and final electron states. If the $ns$ orbitals are not orthonormal to a good numerical precision, the respective integral is not numerically stable, leading to spurious errors, generally of a few per cent. This problem does not arise in the present RPA, MPBT3, and all-order calculations since we do all computations with the orthonormalized basis set wave functions. However, it explains the difference with the DF and RPA results of \cite{GosDzuFla13}, which used DF functions in the RPA calculations. This issue is a potential source of the drastic difference or our RPA values with \cite{GosDzuFla13} for the Fr-like Ac and Th ions. The M1 matrix elements for Fr-like Ac and Th ions are not expected to be significantly different from the Fr and Fr-like Ra values and present RPA and final MBPT3 values for Fr-like isoelectronic sequence shows smooth changes.

The second problem with the
 calculation of the  $s-s$ transitions is a strong cancelation of the large Brueckner-orbital (BO) and structure radiation (SR) corrections.
In the all-order case, the  BO-type term c  given by Eq.(\ref{termc}) is very large but is strongly canceled by SR-type term
\begin{equation}
 Z^{(p)} =   \sum_{mnra} z_{mn} \tilde{\rho}^*_{rmwa} \tilde{\rho}_{rnva}.
\end{equation}
Either of these  terms is at least an order of magnitude larger than the RPA. This issue makes the all-order computation of the M1 $s-s$ matrix elements
 unreliable in its current implementation. Most likely, omission of the triple- and higher-excitations leads to incomplete cancelations and
full inclusion of the other high-order corrections, such as those from non-perturbative triple excitations, non-liner terms, and others
is needed. An inclusion of the perturbative triples or scaling exacerbates the problem instead of correcting it, since
they directly affect only BO-type terms but not the structural radiation.
The $s-d$ transitions do not present such problems: we find some significant contributions from the non-RPA terms described above but no strong cancelations.

To improve upon the RPA results for the $s-s$ transition, we use a form-independent third-order MBPT method introduced in \cite{SavJoh00,SavJoh00a}.
This approach yields electric-dipole transition amplitudes that are equal in the length and velocity forms for transitions in atoms with one valence electron within the framework of relativistic many-body perturbation theory starting from the Dirac-Hartree-Fock approximation.
 For the M1 transitions, where the matrix elements are in velocity form,
such an approach appears to provide more accurate cancelations of the large BO and SR many-body  corrections. Even with the strong cancelations, the remaining corrections are still significant for the
$s-s$ M1 matrix elements. Further improvement of the theoretical accuracy may be achieved with the
development of the form-independent all-order approach.

\section{Conclusion}
In summary, we carried out a systematic
relativistic study of $s-s$ and $s-d$
M1 transitions  in Fr and Fr-like ions,  Cs, Ba$^+$, Rb, Fr, Cs, Ba$^+$, Rb, SR$^+$,
and Ca$^+$ atomic systems. Benchmark comparisons of the $nd$ lifetimes are carried out.
Relativistic, correlation, Breit and negative-energy contributions are
studied. The estimated accuracy of the theoretical $s-d$ E2 matrix elements is very high, better than 1\%.
and is good for the $s-d$ M1 matrix elements. A rough estimate of the accuracy of $s-d$ M1 matrix elements
is given by the difference between the SD and MBPT3 values listed
in Table~\ref{tab-neg} which can exceed 10\%.
We find that inclusion of the correction beyond RPA is essential for accurate calculations of the
M1 matrix elements considered in this work. New high-precision experimental results are urgently needed for the M1 transitions to test theoretical predictions.

 \acknowledgments We thank V.A. Dzuba for helpful discussions.
 This work is partly supported by NSF grant
No. \ PHY-1620687.


\end{document}